\documentclass[aps,prl,twocolumn,showpacs,groupedaddress]{revtex4}  
\usepackage{graphicx}  
\usepackage{dcolumn}   
\usepackage{bm}        
\usepackage{amssymb}   

\begin{document}


\title{A single-ion nonlinear mechanical oscillator}
\author{N.~Akerman}
\author{S.~Kotler}
\author{Y.~Glickman}
\author{Y.~Dallal}
\author{A.~Keselman}
\author{R.~Ozeri}
\affiliation{Physics of Complex Systems, Weizmann Institute of
Science, Rehovot 76100, Israel}

\begin{abstract}
We study the steady state motion of a single trapped ion oscillator
driven to the nonlinear regime. Damping is achieved via Doppler
laser-cooling. The ion motion is found to be well described by the
Duffing oscillator model with an additional nonlinear damping term.
We demonstrate a unique ability of tuning both the linear as well as
the nonlinear damping coefficients by controlling the cooling laser
parameters. Our observations open a way for the investigation of
nonlinear dynamics on the quantum-to-classical interface as well as
mechanical noise squeezing in laser-cooling dynamics.
\end{abstract}


\maketitle


Nonlinear dynamics prevails in many dynamical systems in nature,
introducing a rich behavior such as criticality, bifurcations and
chaos. Nonlinear dynamics on the microscopic scale is especially
interesting as it can shed light on the quantum-to-classical
transition as well as provide a mean to suppress thermal and quantum
noise.

All mechanical oscillators will show nonlinearity when driven far
enough from equilibrium. The simplest such nonlinear oscillator is
the Duffing oscillator which includes a cubic term in the restoring
force \cite{book0}. Recently, such Duffing nonlinear dynamics has
been extensively studied with nano-electromechanical beam
resonators. The basins of attraction of a nano beam oscillator were
mapped \cite{basins_of_attraction}. Noise squeezing and stochastic
resonances were observed close to the Duffing instability
\cite{noise_squeezing, stochastic resonance}. Noise squeezing was
predicted to enable mass and force detection with precision below
the standard thermal limit \cite{Yurke1995, Precision mass
detection, Noise_Enabled_Precision} and possibly below the standard
quantum limit when operating close to the oscillator ground state
\cite{Milburn2008}.
\begin{figure}
\includegraphics[width=8.0 cm]{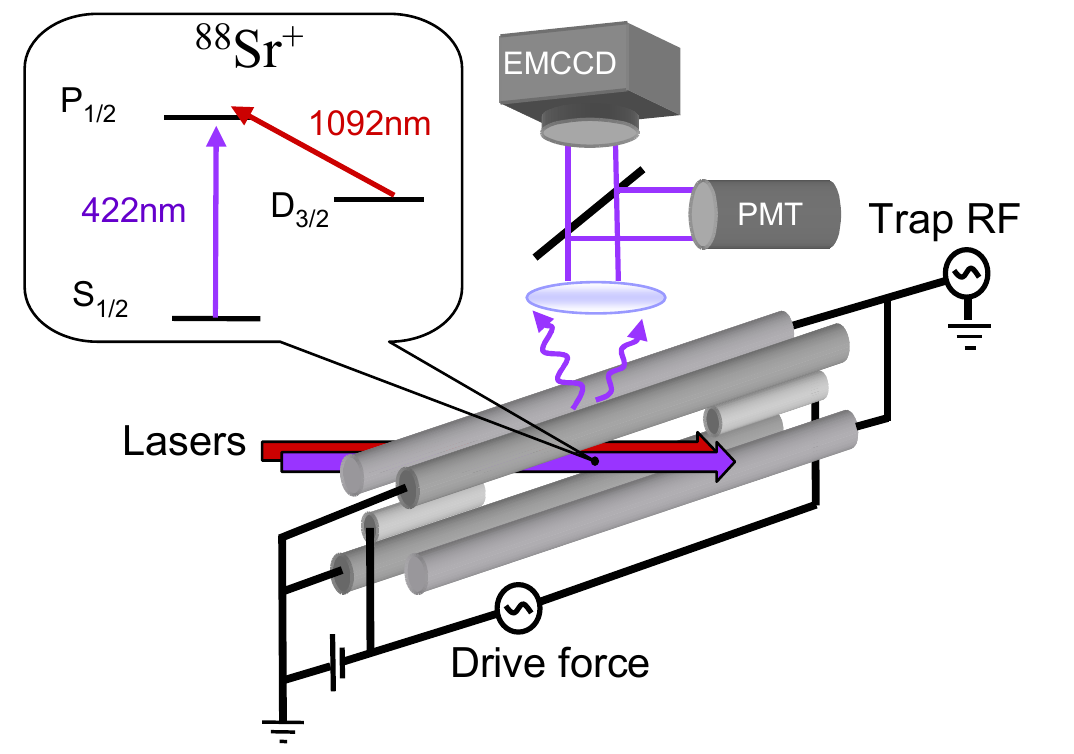}
\caption{\label{fig:epsart} Schematic diagram of the experimental
set-up and the relevant energy levels of $^{88}$Sr$^{+}$ ion. The
positively biased trap end-caps produce a static trapping potential
in the axial direction with a small anharmonicity. The ion
oscillator is driven to the nonlinear regime by a small oscillating
voltage on one of the trap end-caps. Violet and infra-red laser
beams provide laser cooling and optical pumping. Scattered violet
photons are collected by an imaging system and directed either to an
EMCCD camera or to a PMT.}\label{system}
\end{figure}

The mechanical motion of trapped ions is highly controllable and can
be efficiently laser-cooled to the quantum ground state \cite{Didi
RMP 2003}. High fidelity production of Fock, squeezed, and
Schr\"{o}dinger-cat states was demonstrated with a single
trapped-ion \cite{Nonclassical ion states, ion cat state}. At the
temperature range obtained with laser-cooling techniques, quadruple
RF Paul traps are excellently approximated as harmonic. Nonlinearity
in ion motion was observed when several ions are trapped due to
their mutual Coulomb repulsion. Here nonlinearity couples between
the ion-crystal normal modes, even at the single quantum level
\cite{Phonon-Phonon interactions, Nonlinear_coupling}. Trap
nonlinearities are important in the context of resonance ejection in
high resolution mass spectrometry \cite{resonance_ejection_theory}.
However, in these experiments ions are typically not laser-cooled
and furthermore the effect of Coulomb nonlinearities in the large
ion cloud is intertwined with that of the trap. Recently,
amplification saturation of a single-ion ``phonon laser'', resulting
from optical forces that are nonlinear in the ions velocity, was
demonstrated \cite{phononlaser}.

Here, we study the nonlinear mechanical response of a single
laser-cooled $^{88}$Sr$^{+}$ ion, in a linear RF-Paul trap. The
nonlinearity originates from the higher than quadrupolar order terms
in the trapping potential. We find that the ion steady state
response is well described by the Duffing model with an additional
nonlinear damping term \cite{nonlinear damping}. Unlike other
realizations of nonlinear mechanical oscillators, both the linear
and the nonlinear damping components can be precisely controlled.


Our trap has the canonical linear four rods and two end-caps
configuration shown in Fig. \ref{system}. The distance of the ion to
the end-caps and rod-electrodes is $0.65$ mm and $0.27$ mm
respectively. Here we examine only the motion along the axial
direction of the trap. In this direction, trapping is dominated by
the static electric potential due to a positive constant voltage on
the trap end-caps. This potential is well approximated to be
harmonic with $\omega_0/2\pi=438$ KHz. However, as the trap end-caps
do not satisfy the pure electric quadruple boundary condition, a
small octupolar contribution to the electric field results a
positive cubic term in the restoring force and an energy level
difference of $\hbar\omega_0+\hbar\Delta_{nl}n$ where $n$ is the
harmonic oscillator quantum number and $\Delta_{nl}/2\pi=0.8$ mHz is
the nonlinear dispersion. This nonlinearity becomes increasingly
important with growing oscillation amplitude. The ion is driven to
the nonlinear regime by adding a small oscillating voltage to one of
the trap end-caps. The ion is Doppler-cooled by scattering photons
from a single laser beam, slightly red-detuned from the
$S_{1/2}\rightarrow P_{1/2}$ transition at $422$ nm. To prevent
population accumulation in the $D_{3/2}$ meta-stable level we repump
the ion on the $D_{3/2}\rightarrow P_{1/2}$ transition at $1092$ nm.

We measure the steady-state oscillation amplitude of the ion as we
slowly scan the drive frequency, $\omega$, across the harmonic
resonance, $\omega_0$. The scan is from lower to higher frequencies
(positive sweep) or vice versa (negative sweep). Photons scattered
during the cooling process are collected by an imaging system (N.A.
= 0.31), and are either directed towards an Electron-Multiplying CCD
(EMCCD) camera or a Photo Multiplier Tube (PMT).

\begin{figure}
\includegraphics[width=8.5 cm]{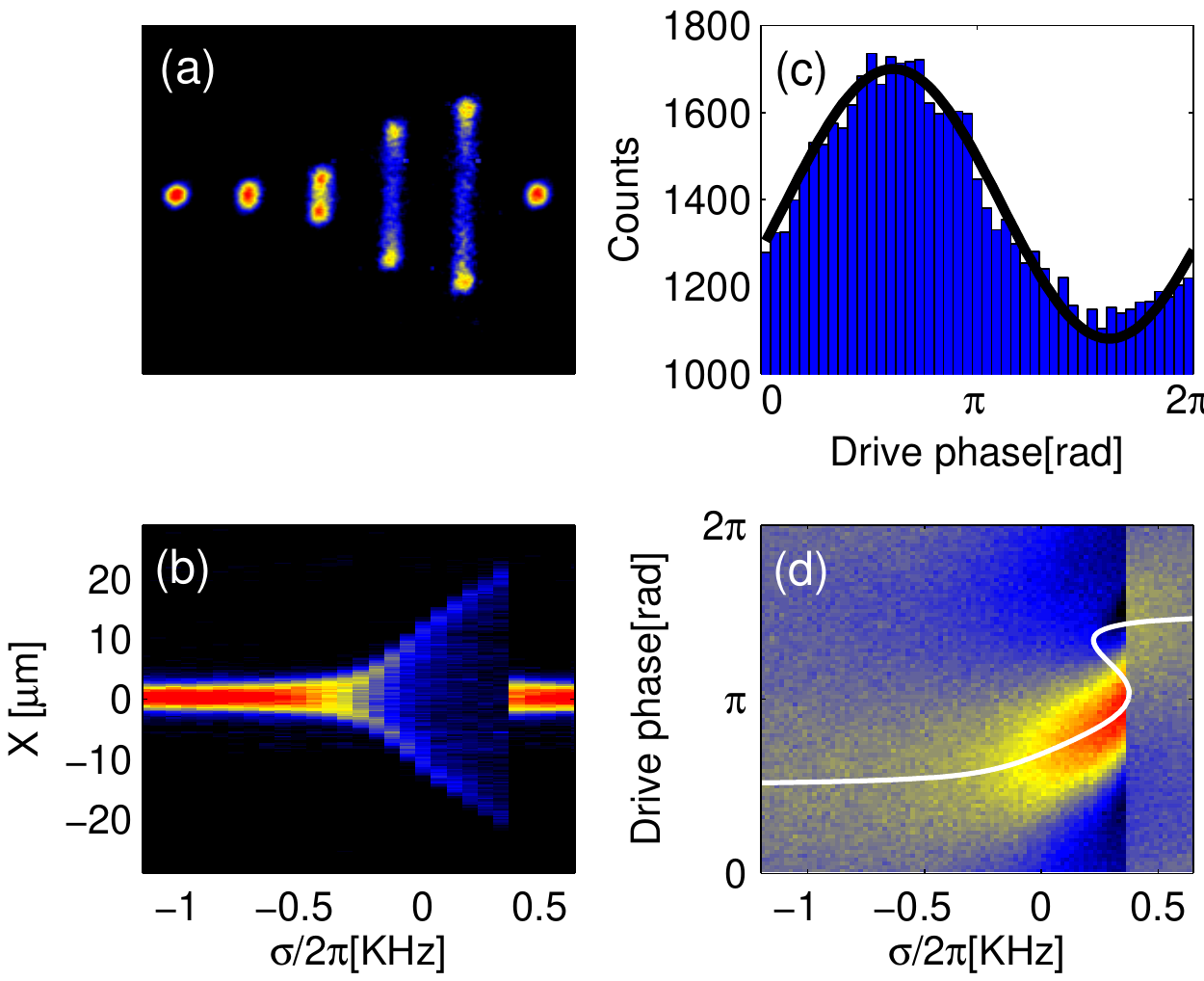}
\caption{\label{fig:epsart} Driven ion-oscillator amplitude and
phase. (a) Time-averaged ion images taken at various drive
frequencies. (b) Columns are time-averaged images, integrated along
the direction perpendicular to ion-motion, during a positive
frequency scan. (c) A histogram of the number of photons detected at
different driving force phases. (d) Columns are photon phase
histograms taken during a positive frequency scan. The solid line is
the theoretical phase given by Eq.\ref{eq:DuffingPhase} shifted by a
constant to match the peak in the histograms.}\label{Ion motion}
\end{figure}

We measure the amplitude of motion by taking time-averaged images of
the ion as shown in Fig.\ref{Ion motion}(a). The image is then
integrated along the direction perpendicular to motion to produce a
single curve. Every column in Fig. \ref{Ion motion}(b) corresponds
to a curve produced this way, for a positive frequency sweep. As
seen, the oscillation amplitude increases as the drive frequency
approaches $\omega_0$, continues to increase passed $\omega_0$,
until at a given critical drive frequency, $\omega_m$, abruptly
collapses to a significantly lower value. We extract the ion
oscillation amplitude by fitting the curve to the expected
time-averaged position distribution. The blue and red lines in Fig.
\ref{frequency response} are the measured amplitudes for positive
and negative sweeps respectively. Different curves correspond to
different drive amplitudes. The asymmetry and hysteresis as well as
the abrupt amplitude changes at specific critical drive frequencies
are a clear deviation from the driven harmonic oscillator response.
As expected from a positive nonlinearity, the oscillator
self-frequency is ``pulled'' to higher values at higher oscillation
amplitudes.

In order to measure the phase difference between the ion-oscillator
and the driving force, we time-stamp each photon measured by the PMT
within a single drive period to allocate it with a corresponding
drive phase. The instantaneous photon scattering rate from the
cooling laser beam is determined by the ions' instantaneous velocity
through its associated Doppler shift \cite{scattering rate}. A
histogram of the measured photon phases is shown in Fig.\ref{Ion
motion}(c). A clear sinusoidal oscillation of the photon scattering
rate yields the ion-oscillator phase. The columns in Fig.\ref{Ion
motion}(d) are photon phase histograms for a positive drive
frequency sweep. As seen, at the critical frequency, $\omega_m$, a
phase jump of $1.2$ radians in the oscillator motion accompanies the
sudden change in oscillation amplitude.

Our observations are well accounted for by the Duffing oscillator
model. The Duffing equation of motion is,
\begin{equation}
\ddot{x} + 2\mu\dot{x} + \omega_{0}^{2}x + \alpha x^3 = k
\cos(\omega t) \label{eq:one}.
\end{equation}
Here $x$ is the displacement of the ion from its equilibrium
position, $\alpha$ is the an-harmonic coefficient, $\mu$ is the
linear damping coefficient and $k$ is the drive amplitude. The
recoil noise inherent to the spontaneous photon scattering process,
which would appear as a Langevin force term, is neglected. An
approximate solution to Eq.\ref{eq:one} can be obtained by the
multiple scale method \cite{book0}. Here the solution has the from
$x(t)=a(t)\cos(\omega t - \phi)$ , where $a(t)$ is a slowly-varying
oscillation amplitude and $\phi$ is the oscillator phase. The
steady-state solution for $a$ solves,
\begin{equation}
\sigma=\frac{3\alpha}{8\omega_0}a^2\pm\sqrt{\frac{k^2}{4\omega_{0}^{2}a^2}-\mu^2}
\label{eq:DuffingAmp},
\end{equation}
where $\sigma=\omega-\omega_0$ is the drive detuning. The
steady-state solution for $a$, at a fixed $k$, vs. drive frequency
is shown by the black line in the inset of Fig.\ref{frequency
response}. Above a critical amplitude $a_c$, $a$ trifurcates into
three solutions. One solution with small and one with large
amplitude, are stable, while the third, intermediate amplitude
solution, is unstable and is positioned on the state-space
separatrix. This bistablity persists until the high amplitude
solution reaches a maximal value, $a_{m}$, at which the drive force
is overwhelmed by damping and the oscillator is forced into a single
stable solution. Positive and negative frequency scans carry the
oscillator into the bistability region along different stable
attractors leading to the observed hysteresis as illustrated by the
arrows in the inset. To compare with our data we independently
measure all the parameters in Eq.\ref{eq:one}. The driving force
amplitude, $k$, is measured by observing ion displacement vs.
end-cap voltage, $\omega_0$ is measured via ion response in the
linear regime. A value of $\alpha/4\pi^2=1.24\pm0.03\cdot10^{18}$
$Hz^2/m^2$ is measured using the observed dependence of $a_{m}$ on
$\sigma_m = \omega_m - \omega_0$, the instability detuning,
$a_{m}=\sqrt{8\omega_0\sigma_m/3\alpha}$. A value of
$\mu/2\pi=39.2\pm0.3$ Hz, which result in a quality factor $Q=5590$,
is evaluated using the variation of $a_m$ with the drive amplitude,
$a_{m}=k/(2\mu\omega_0)$. The blue and red circles in the inset are
the measured amplitudes, for positive and negative scans
respectively, showing good agreement with the theoretical curve.
\begin{figure}[htb]
\includegraphics[width=8.5 cm]{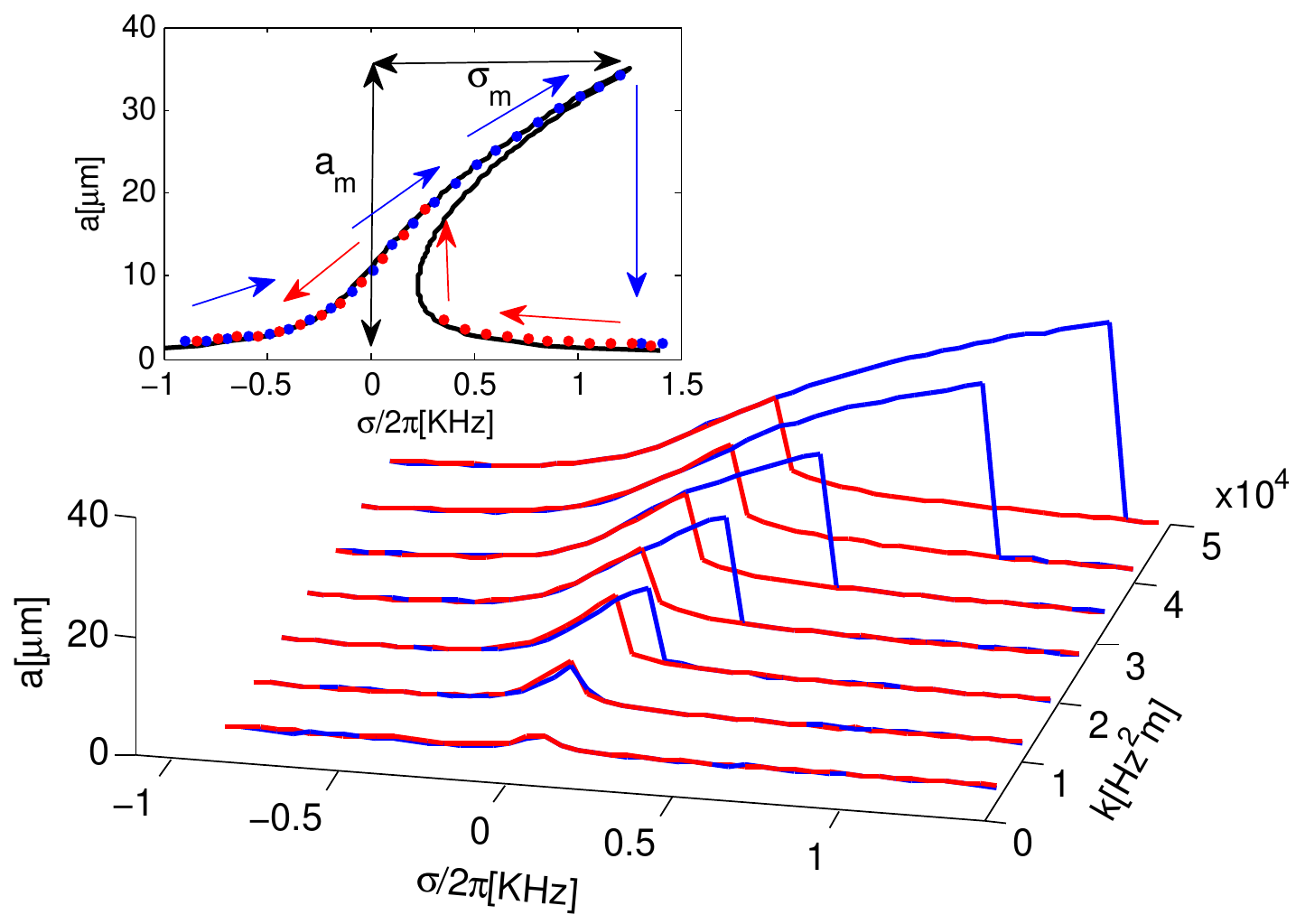}
\caption{Measured oscillator amplitude vs. drive frequency for
various driving force amplitudes and both positive (blue) and
negative (red) scans. The inset shows the Duffing model calculation
(black solid line) and our measured amplitudes (blue circles -
positive scan; red circles - negative scan)}\label{frequency
response}
\end{figure}

The Duffing oscillator steady state phase is given by,
\begin{equation}
\tan(\phi)=\frac{8\mu\omega_0}{3\alpha a^2 - \omega_0\sigma}
\label{eq:DuffingPhase}.
\end{equation}
The white solid line in Fig.\ref{Ion motion}(d) shows the
theoretical phase curve vs. drive frequency for our experimental
parameters, showing good agreement with our data.

Linear damping is a very good approximation for most mechanical
oscillators, as typically dissipation originates from coupling of
the oscillator to an ohmic bath. Recently, the contribution of
non-linear damping to the motion of a nano beam resonator was
studied \cite{Buks nonlinear_damping}. In our experiment damping
results from the change in radiation pressure vs. ion velocity. When
the laser frequency is tuned below the cooling transition, the
leading contribution is indeed linear in the ions' velocity.
However, as the oscillation amplitude increases or the cooling-laser
detuning reduced, the effect of damping force terms that are
nonlinear in the ions' velocity increases \cite{scattering rate}.

\begin{figure}[htb]
\includegraphics[width=8.5 cm]{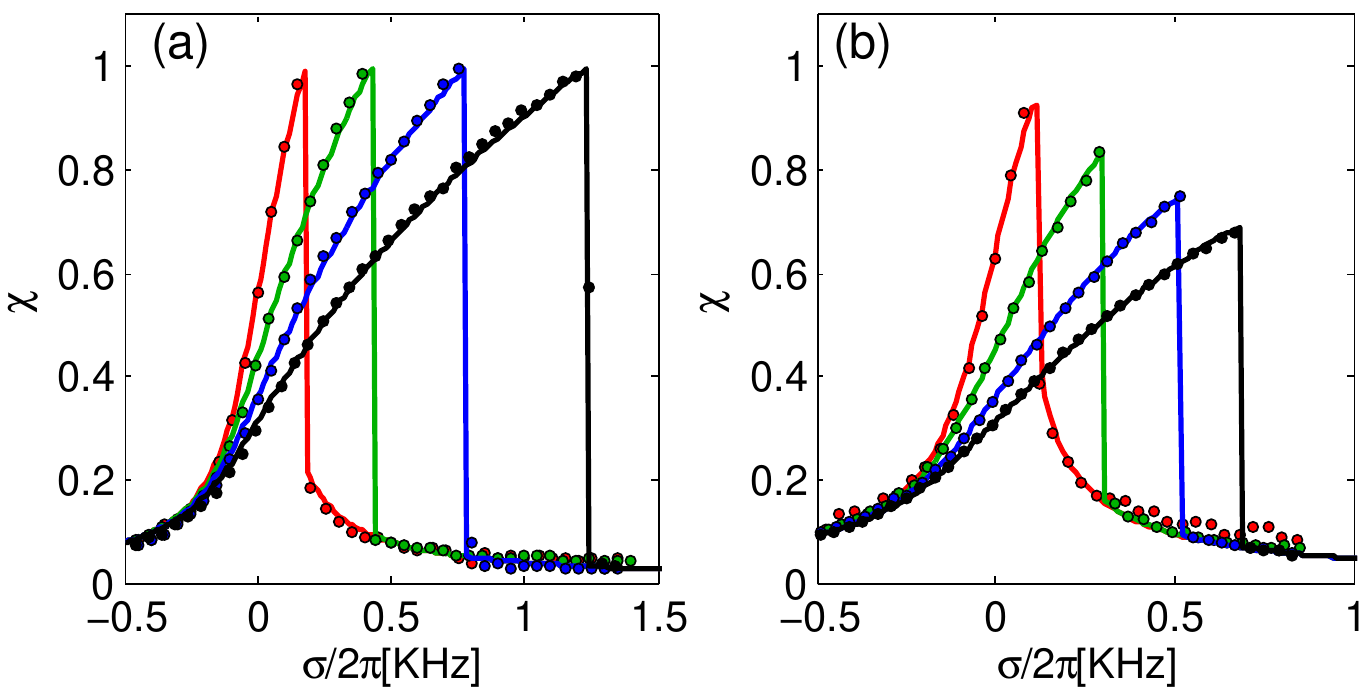}
\caption{\label{fig:epsart} Calculated and measured responsivity,
$\chi = 2\mu\omega_0 a/k$, for positive drive scans. Different
curves correspond to different drive amplitudes. Due to small drifts
in $\omega_0$ ($<$100Hz) each curve was separately shifted on the
frequency axis to fit the theoretical curve. (a) Linear damping, the
cooling laser detuning $\delta/2\pi=-420$ MHz and $\gamma= 0$. The
maximal responsivity is independent of drive amplitude $k$. (b)
Nonlinear damping, $\delta/2\pi=-160$ MHz and
$\omega_0^2\gamma/2\pi=0.09$ $\mu$ m$^{-2}$Hz. The maximal
responsivity decreases as $k$ increases.}\label{responsivity}
\end{figure}

To account for nonlinear damping, Eq.\ref{eq:one} is modified to
include a term which is cubic in the oscillator velocity,
\begin{equation}
\ddot{x} + 2\mu\dot{x}+\gamma\dot{x}^{3} + \omega_{0}^{2}x + \alpha
x^{3} = k \cos(\omega t). \label{eq:three}
\end{equation}
Here $\gamma$ is the cubic damping coefficient. The steady-state
amplitude is now a solution of \cite{Buks
nonlinear_damping,Lifshitz2008},
\begin{eqnarray}
& \frac{9}{16}(\alpha^2+\gamma^2\omega_0^6)a^6 &+
3\omega_0(\mu\gamma\omega_0^3-\sigma\alpha)a^4 \nonumber\\
&& + 4\omega_0^2(\sigma^2 +\mu^2)a^2 - k^2 = 0.\label{nonliear
dissipation}
\end{eqnarray}
When $\gamma>0$, nonlinear damping acts to effectively increase
dissipation for larger oscillation amplitudes. Unlike the linear
damping case, $a_m$ does not increase linearly with $k$ but is
rather limited by the growing dissipation. We find $\mu$ and
$\gamma$ by a maximum likelihood fit of the measured $a_m$ vs. $k$
curve to the solution of Eq. \ref{nonliear dissipation}. It is
instructive to look at the responsivity, $\chi=2\mu\omega_0a/k$, in
order to distinguish linear from nonlinear damping
\cite{Lifshitz2008}. In Fig.\ref{responsivity} we plot the measured
$\chi$ for positive scans and various drive amplitudes, $k$, for two
different cooling-laser detuning values, $\delta$. In
Fig.\ref{responsivity}(a) $\delta/2\pi=-420$ MHz, $\gamma=0$ and the
maximal responsivity is seen to be independent of $k$. In
Fig.\ref{responsivity}(b) $\delta/2\pi=-160$ MHz,
$\omega_0^2\gamma/2\pi=0.09\pm0.002$ $\mu$m$^{-2}$Hz and the maximal
responsivity decreases as $k$ increases. The linear dissipation
term, $\mu$, is similar in both cases. The solid lines are the
solutions of Eq.\ref{nonliear dissipation} showing good agreement
with the data.

\begin{figure}[htb]
\includegraphics[width=8.5 cm]{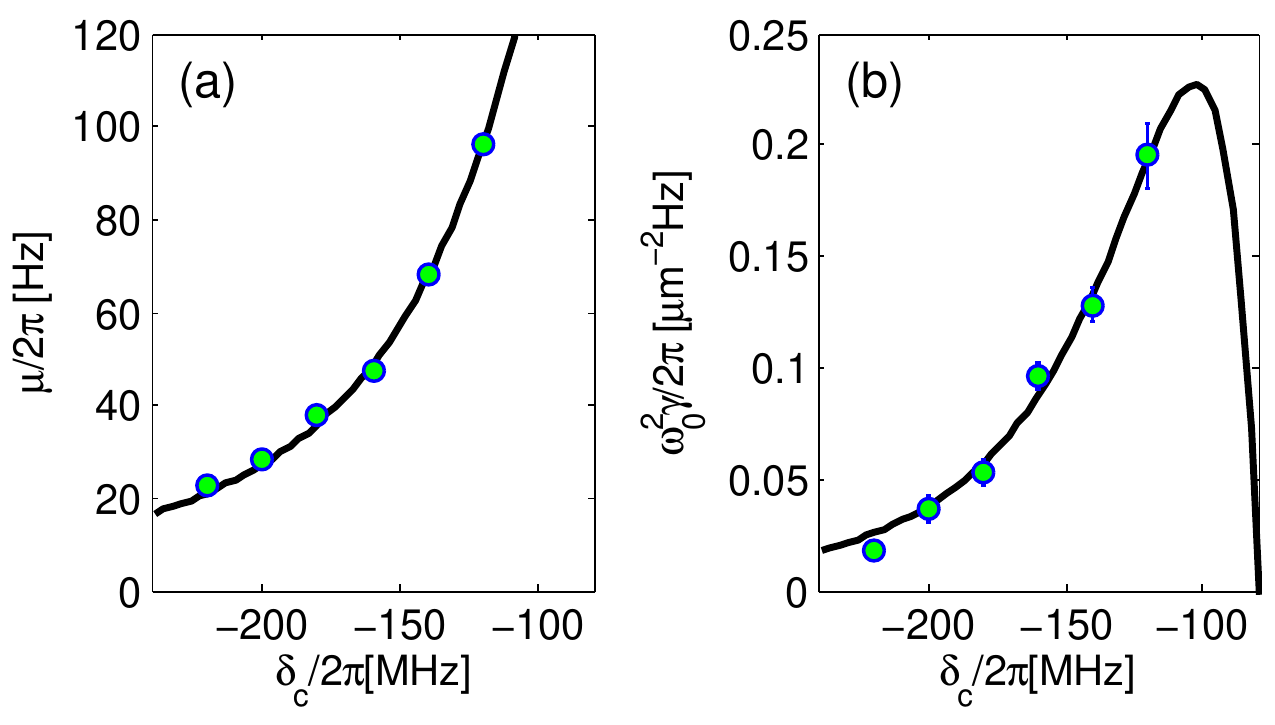}
\caption{\label{mu and gamma} (a) Linear and (b) cubic damping
coefficients for various cooling-beam detunings. Filled circles are
measured values and solid lines are calculated using equations
\ref{scattering force} and \ref{damping coeffs}.}\label{dissipation}
\end{figure}

We next repeat the measurement of $\mu$ and $\gamma$ for various
cooling-laser detunings at a fixed repump-laser frequency and lasers
intensities. The measured $\mu$ and $\gamma$ vs. $\delta$ are shown
in figures \ref{dissipation}(a) and \ref{dissipation}(b)
respectively. To compare with the theoretically predicted values we
write the cooling-laser scattering force,
\begin{equation}
F_s(\dot x)= \hbar k_{c}\Gamma\rho_{p}(\delta_c+k_c\dot
x,\delta_r+k_r\dot x),\label{scattering force}
\end{equation}
where $k_{c/r}$ and $\delta_{c/r}$ are the wave-vectors and
detunings of the cooling and repump lasers respectively,
$\Gamma=2\pi\times 21$ MHz is the spectral linewidth of the
$P_{1/2}$ level and $\rho_p$ is the $P_{1/2}$ population. The
damping coefficients are therefore given by the appropriate
derivatives,
\begin{equation}
\mu=\frac{1}{2 m} \frac{dF_{s}}{d\dot{x}} ~~;~~ \gamma=\frac{1}{6
m}\frac{d^3F_{s}}{d\dot{x}^3}.\label{damping coeffs}
\end{equation}
Here $m$ is the ion mass. We calculate $\rho_P$ by numerically
solving the eight coupled Bloch equations, corresponding to the
population in all states in the $S_{1/2}$, $P_{1/2}$ and $D_{3/2}$
levels coupled by the cooling and repump lasers. The cubic damping
coefficient is highly sensitive to different laser parameters due to
the presence of dark resonances. The solid lines in figures
\ref{dissipation}(a) and \ref{dissipation}(b) are the calculated
$\mu$ and $\gamma$ showing good agreement with our measured values.
The two lasers intensities and the repump-laser detuning were used
as fit parameters, yielding values that agree within 20\% with their
measured value.

An additional nonlinear damping term, proportional to $x^2\dot{x}$,
results from the laser beam finite size ($100$ $\mu$m FWHM) and has
an identical effect to that of $\gamma$ on the steady-state motion
\cite{Lifshitz2008, nonlinear damping}. This term was calculated to
be small relative to $\gamma$ \cite{gamma remark} and was taken into
account in Fig.\ref{mu and gamma}.

In conclusion, we have driven a single-ion oscillator to the
nonlinear regime. The ion steady-state motion, showing a bifurcation
into two stable attractors and hysteresis, is well described by the
Duffing oscillator model with an additional nonlinear damping term.
Unlike previously studied nonlinear mechanical oscillators, here
both the linear and nonlinear parts of dissipation can be tuned with
the cooling laser parameters.

The study of the nonlinear motion of trapped laser-cooled ions opens
several exciting research avenues. Since trapped atomic-ions can be
cooled to the quantum ground state, they are an excellent platform
to study nonlinear behavior in the quantum regime. As shown in
\cite{classical_quantum_transition}, unlike the simple harmonic
oscillator, a Duffing oscillator will demonstrate a clear
quantum-to-classical transition even when classically driven.
Moreover, as the ion-spin can be entangled with its motion, it will
be possible to form a coherent superposition of the two attractors
states of motion. Laser-cooling of a nonlinear driven ion-oscillator
has several interesting aspects that can be further explored. Since
the Doppler shifts associated with the oscillation amplitudes in the
nonlinear regime are significant compared with the cooling
transition line-width, the laser-cooling force is largely nonlinear
in the oscillator velocity. Furthermore, the thermal state generated
by laser-cooling is the result of balance between the damping force
and the inherent heating due to the recoil noise from spontaneous
photon scattering. Close to the Duffing instability, the
ion-oscillator response to noise is quadrature dependent. One noise
quadrature is largely enhanced whereas the other quadrature is
suppressed \cite{noise_squeezing}. Laser-cooling in this case is
likely to produce squeezed states of motion.

This work was partially supported by the ISF Morasha program and the
Minerva foundation.


\begin{thebibliography}{20}
\bibitem{scattering rate}
As the life-time of the P$_{1/2}$ level ($8$ ns) is much shorter
than the oscillation period ($228$ ns), the photon scattering rate
instantaneously adjusts as the ions' velocity varies.
\bibitem{gamma remark}
The ratio between the two terms depends on the cooling laser
detuning and is always below $0.2$.
\end{thebibliography}
\end{document}